\documentclass{kapproc} % Computer Modern font calls

\usepackage{procps} 

\usepackage[dvips]{graphicx}

\upperandlowercase

\kluwerbib % will produce this kind of bibliography entry:

\let\lcitebracket(
\let\rcitebracket)

\newcommand{\oml}{\Omega_{\Lambda}}
\newcommand{\omm}{\Omega_{m}}

\newcommand{\gr}{\kern 2pt\hbox{}^\circ{\kern -2pt K}} %  ====> GRADI KELVIN
\newcommand{\brr}{\begin{array}}
\newcommand{\err}{\end{array}}
\newcommand{\ltsima}{$\; \buildrel < \over \sim \;$}
\newcommand{\simlt}{\lower.5ex\hbox{\ltsima}}
\newcommand{\gtsima}{$\; \buildrel > \over \sim \;$}
\newcommand{\simgt}{\lower.5ex\hbox{\gtsima}}
\begin{document}

\articletitle[ X-ray properties] 
 { X-ray Cluster Properties \\
in SPH Simulations of Galaxy Clusters}

\author{R. Valdarnini,\altaffilmark{1} }

\affil{\altaffilmark{1}SISSA, Vie Beirut 2-4 34014, Trieste Italy}

\begin{abstract}
 Results from a large set of hydrodynamical SPH simulations
of galaxy clusters in a flat LCDM cosmology are used to investigate 
cluster X-ray properties. The physical modeling of the gas includes radiative 
cooling, star formation, energy feedback and metal enrichment that
follows from the explosions of SNe type II and Ia.
The metallicity dependence of the cooling function is also taken into account.
It is found that the luminosity-temperature relation of simulated clusters 
is in good agreement with the data, and the X-ray properties of cool 
clusters are unaffected by the amount of feedback energy that has heated the 
intracluster medium (ICM).  The fraction of hot gas $f_g$ at the virial radius 
increases with $T_X$ 
and the distribution obtained from the simulated cluster sample is consistent 
  with the observational ranges.

\end{abstract}


\section{Introduction}
There is a wide observational evidence
\cite{ala98,mar98}
that the observed cluster X-ray luminosity scales with temperature with a slope
which is  steeper than that predicted by the self-similar scaling relations
($L_X \propto T_X^3$).
 This implies that low temperature clusters have central densities lower
than expected  \cite{pon99}.
 This break of self-similarity is usually taken as a strong evidence that
non-gravitational heating of the ICM has played an important role in the ICM
evolution.
A popular model which has been considered as a heating
mechanism for the ICM is supernovae (SNe) driven-winds \cite{whi91}.
An alternative view is that radiative
cooling and the subsequent galaxy formation 
 can explain the observed $L_X-T_X$ relation because of the removal of 
low-entropy gas at the cluster cores \cite{br00}.             
In this proceedings I present preliminary results from a large set of 
hydrodynamical SPH simulations of galaxy clusters. The physical modeling of 
the gas includes a number of processes (see later), and the simulations have 
been performed in order to investigate the consistency of simulated cluster 
scaling relations  against a number of data.

\section{Simulations and results }
 Hydrodynamical TREESPH simulations 
have been performed in physical coordinates for a sample of 120 test clusters.  
A detailed description of the  procedure can be found in  
Valdarnini (2003, V03).
The cosmological model is a flat CDM model, with vacuum energy density
$\oml=0.7$, matter density parameter $\omm=0.3$ and Hubble constant $h=0.7
=H_0/100 Km sec^{-1} Mpc^{-1}$. $\Omega_b=0.019h^{-2}$ is the value of 
the baryonic density. 
The clusters are the 120 most massive ones identified at $z=0$ in a 
cosmological 
$N-$body run with a box size of $200 h^{-1}$ Mpc.
 The virial temperatures range from 
$\sim 6 KeV$  down to $\sim 1KeV$.
The simulations have  $N_g \simeq 70,000$ gas particles. 
The cooling function of the gas particles also depends on the gas metallicity, 
and cold gas particles are subject to star formation.
Once a star particle is created it will release energy into the
 surrounding gas through SN explosions of type II and Ia.
The feedback energy ($10^{51}$ erg) is returned to the 
nearest neighbor  gas particles, according to its
lifetime and IMF. SN explosions also inject enriched material into the ICM,
thus increasing its metallicity with time. 

Observational variables of the simulated clusters are plotted in Fig. 1
as a function of the cluster temperature $T_X$. Different symbols are for
 different redshifts ($z=0$, $z=0.06$ and $z=0.11$).
 For the sake of clarity, not all the points of the numerical sample are 
plotted in the figure.
Global values $A_{Fe}=M_{Fe}/M_H$ of the iron abundances for the simulated 
clusters are shown in panel (a). A comparison with real data of a small
sample subset (four clusters , see V03) shows a good agreement with 
the measured values. 
The values of $A_{Fe}$  appear to increase with $T_X$, even though
the observational evidence of an iron abundance increasing with $T_X$
is statistically weak \cite{mu97}.    

\begin{figure}[!h]
\includegraphics[width=12truecm,height=10truecm]{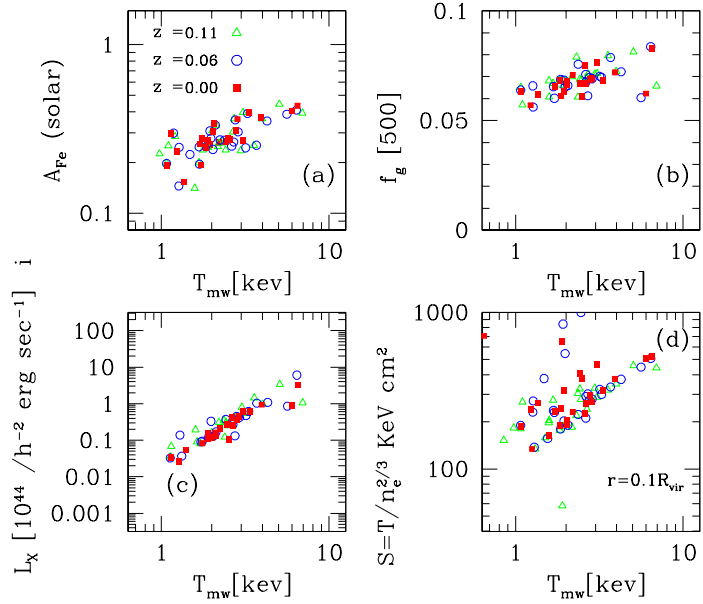}
\caption{
{\bf (a)}: 
Average iron abundances of the simulated cluster sample at $r=0.5h^{-1}Mpc$
are plotted versus cluster temperatures. Different symbols refer to different 
redshitfs.
{\bf (b)}: As in panel (a), but for the 
gas fractions $f_g=M_g(<r)/M_T(<r)$; the fractions are evaluated at the 
radius within which $\rho/\rho_c=\delta$, with $\delta=500$. 
{\bf (c)}: 
Bolometric X-ray luminosities are plotted as a function of the temperature.
{\bf (d)}: Cluster core entropies 
 $S(r)= k_B T(r)/n_e(r)$ at the radius $r=0.1r_{vir}$
are shown against cluster temperatures.
}
\label{fig:prof}
\end{figure}

The fraction of hot gas $f_g=M_g/M_T$ is defined within a given radius as the
 ratio of the mass of hot gas $M_g$ to the total cluster mass $M_T$. 
The fraction $f_g$ has been calculated  at a radius enclosing a gas overdensity
of $\delta=500$ relative to the critical density.
The values of $f_g$ of the numerical cluster sample are plotted in 
panel (b). 
There is a clear tendency for the $f_g$ distributions to increase with
cluster temperature. This is in accord with theoretical predictions of the
 radiative cooling model \cite{br00}, and with numerical 
simulations \cite{mu02,da02}. A subsample of the $f_g$ distribution at 
$z=0$ is found to be statistically consistent (V03) with the data of 
Arnoud \& Evrard (1999). Furthermore, Sanderson et al. (2003) found
 strong observational evidence for a dependence of $f_g$ with $T_X$.

The values of the bolometric X-ray luminosity $L_X$ are shown in 
panel (c) as a function of the cluster temperature. Mass-weighted temperatures
have been used as unbiased estimators of the spectral temperatures
\cite{ma01}.
A central region of size $50 h^{-1} Kpc$ has been excised ~\cite{mar98} in 
order to remove the contribution to $L_X$ of the central cooling flow.
  The $L_X$ of the simulations at $z=0$ are in good
agreement with the data~(V03) over the entire range of temperatures.
An additional run has been performed for the cluster with the lowest
temperature, but with a SN feedback energy of $10^{50}$ erg for both SNII and 
Ia.  The final $L_X$  of the run 
 is very similar to that of standard run. This demonstrates that the
final X-ray luminosities of the simulations are not sensitive to the 
amount of SN feedback energy that has heated the ICM. 

The core entropies of the clusters are displayed in panel (d) against the 
 cluster temperatures.
The cluster entropy is defined as $S(r)= k_B T(r)/n_e(r)$, where $n_e$ is
the electron density. The core entropy is calculated at a radius which is 
$10\%$ of the cluster virial radius. 
The result indicates that for low temperatures part of the sample has 
clusters with core entropies which approach a floor at $\sim 100 $kev cm$^{-2}$,
while for some clusters the decline of entropy with temperature is steeper
and close to  the self-similar predictions ($S \propto T$). This different
behaviour could be due to the different dynamical histories of the 
clusters.
It is worth stressing that for all the plotted quantities there is 
little evolution below $z=1$.

To summarize, simulation results give final X-ray properties 
 in broad agreement with the data. 
These findings support the so-called radiative model, where the X-ray 
properties of the ICM are driven by the efficiency of galaxy formation rather
 than by the heating due to non-gravitational processes.

\end{document}